\documentclass[12pt, epsfig]{article}
\usepackage{epsfig}
\usepackage{amssymb}
\usepackage{amsfonts}
\usepackage{bbm}
\newskip\humongous \humongous=0pt plus 1000pt minus 1000pt

\newif\ifdtup

\catcode`\@=11

\@addtoreset{equation}{section}
\def\theequation{\thesection\arabic{equation}}

\def\@normalsize{\@setsize\normalsize{15pt}\xiipt\@xiipt
\abovedisplayskip 14pt plus3pt minus3pt%
\belowdisplayskip \abovedisplayskip
\abovedisplayshortskip \z@ plus3pt%
\belowdisplayshortskip 7pt plus3.5pt minus0pt}

\def\small{\@setsize\small{13.6pt}\xipt\@xipt
\abovedisplayskip 13pt plus3pt minus3pt%
\belowdisplayskip \abovedisplayskip
\abovedisplayshortskip \z@ plus3pt%
\belowdisplayshortskip 7pt plus3.5pt minus0pt
\def\@listi{\parsep 4.5pt plus 2pt minus 1pt
     \itemsep \parsep
     \topsep 9pt plus 3pt minus 3pt}}

\relax

\catcode`@=12

\catcode`\@=11

\def\section{\@startsection{section}{1}{\z@}{3.5ex plus 1ex minus
   .2ex}{2.3ex plus .2ex}{\large\bf}}

\def\thesection{\arabic{section}.}

\def\appendix{\setcounter{section}{0}
 \def\thesection{Appendix \Alph{section}:}
 \def\theequation{\Alph{section}.\arabic{equation}}}

\begin{document}

\newcommand{\be}{\begin{equation}}
\newcommand{\ee}{\end{equation}}
\newcommand{\bqa}{\begin{eqnarray}}
\newcommand{\eea}{\end{eqnarray}}
\newcommand{\beq}{\begin{equation}}
\newcommand{\eeq}{\end{equation}}
\newcommand{\bea}{\begin{eqnarray}}

\newcommand{\non}{\nonumber}
\def\de{\partial}
\def\Tr{ \hbox{\rm Tr}}
\def\const{\hbox {\rm const.}}
\def\im{\hbox{\rm Im}}
\def\re{\hbox{\rm Re}}
\def\Arg{\hbox {\rm Arg}}
\def\Re{\hbox {\rm Re}}
\def\Im{\hbox {\rm Im}}
\def\diag{\hbox{\rm diag}}
\def\ket#1{|{#1}\rangle}
\def\bra#1{\langle {#1} |}
\def\brc{\langle}
\def\ckt{\rangle}
\def\de{\partial}

\begin{titlepage}

{\hfill IFUP-TH/2007-35, DAMTP-2008-4 }  
\bigskip
\bigskip
\bigskip
\bigskip

\begin{center}
{\Large {\bf
Non-Abelian Vortices \\ without 
Dynamical Abelianization
 } }
\end{center}
\vspace{1em}

\begin{center}
{\large   Daniele DORIGONI $^{(1)}$,  Kenichi KONISHI $^{(1,2)}$,   Keisuke OHASHI  $^{(3)}$
 \vskip 0.10cm
 }
\end{center}

\begin{center}
{\it   \footnotesize
Dipartimento di Fisica ``E. Fermi" -- Universit\`a di Pisa $^{(1)}$, \\
Istituto Nazionale di Fisica Nucleare -- Sezione di Pisa $^{(2)}$, \\
     Largo Pontecorvo, 3,  Ed. C, 56127 Pisa,  Italy $^{(1,2)}$ \\
Department of Mathematics and Theoretical Physics (DAMTP) $^{(3)}$ \\
University of Cambridge, 
Cambridge, UK 
   }

\end {center}

\vspace{0.5  em}

\noindent
{\bf Abstract:}
Vortices carrying truly non-Abelian flux moduli, which do not dynamically reduce to Abelian vortices,  are found in the context of softly-broken  ${\cal N}=2$ supersymmetric chromodynamics (SQCD).  By tuning  the bare quark masses appropriately we identify the vacuum in which the underlying $SU(N)$ gauge group is partially broken to $SU(n) \times SU(r) \times U(1)/{\mathbbm Z}_{K}$,  where $K$ is the least common multiple of $(n, r)$, and with $N_{f}^{su(n)}=n$ and  $N_{f}^{su(r)}=r$  flavors of light quark multiplets.  At much lower energies the gauge group is broken completely by the squark VEVs,  and vortices develop which carry non-Abelian flux moduli  $CP^{n-1}\times CP^{r-1}$.  For $n>r$ we argue that the 
$SU(n)$  fluctuations become strongly coupled and Abelianize, while leaving  weakly fluctuating  $SU(r)$ flux moduli. This allows us to   
 recognize the semi-classical origin of the light non-Abelian monopoles found earlier in the fully quantum-mechanical treatment of $4D$ SQCD. 

\begin{flushright}
January  2008

\end{flushright}

\end{titlepage}

\bigskip

\hfill{}
\bigskip

\section{Introduction}

       Attempts to understand better the mechanism of confinement of non-Abelian variety, which is probably the case for the realistic world of QCD,  has eventually led to the discovery of vortices 
 with non-Abelian continuous flux moduli \cite{HT},\cite{ABEKY},   triggering a remarkable development of research activity in related problems \cite{ABEK}-\cite{EHT}.   A typical system considered is a $U(n)$  theory with $N_{f}=n$ scalar quark flavors,  whose vacuum expectation value (VEV) breaks the gauge symmetry completely,  leaving however the color-flavor diagonal $SU(n)_{C+F}$ symmetry unbroken  (color-flavor locking).  Vortices in such a system develop a continuous zeromodes (moduli) parametrizing 
 \[    SU(n) / SU(n-1)\times U(1) \sim CP^{n-1}\;,
 \]
where the divisor represents the symmetry respected by individual vortices. When the vortex orientation is allowed to fluctuate along $z$ (the direction of the vortex length) and in time $t$,  the dynamics of such fluctuations is described by a two dimensional $CP^{n-1}$ sigma model \cite{ABEKY,SY,HT2}.  If the original system is the bosonic sector of a ${\cal N}=2$ supersymmetric model, the sigma model has $(2,2)$ supersymmetry, as half of the supersymmetry is broken by the vortex.   In the infrared limit, the sigma model becomes strongly coupled, and the $2D$ system reproduces exactly \cite{SY,HT2} the dynamics of the corresponding $4D$ gauge
theory in {\it Coulomb phase},  encoded by Seiberg-Witten curves \cite{SW1,SW2,curves}, realizing thus the idea of duality between two-dimensional sigma model and a four-dimensional gauge theory discussed earlier by Dorey \cite{ND}. 

Beautiful as it may be,  the very result of the analysis shows that the vortices considered in \cite{ABEKY,SY,HT2} dynamically Abelianize to Abrikosov-Nielsen-Olesen (ANO) vortices (see the next section).    This fact can be seen both in two and four dimensions.  In the sigma model analysis, the fluctuations inside the vortex become strongly coupled and generates the mass scale, $\Lambda$;  there are $n$ degenerate ground states \cite{GSY} (Witten-CFIV index \cite{Witten,CFIV}).  Monopoles appear as kinks (domain walls) connecting two adjacent vortex ground states.  Each monopole is confined by two 
vortices carrying the ``adjacent''  $U(1)$  fluxes, a typical situation for a monopole arising from the breaking of $SU(2) \subset U(n)$ to $U(1)$.  The global $SU(N_{f})=SU(n)$ {\it flavor} symmetry is not spontaneously broken by the vortex dynamics \footnote{Of course this is consistent with Coleman's theorem.};  this however does not contradict the fact that the monopoles in the infrared carry only Abelian magnetic $U(1)^{n}$ charges.

In four dimensions, the model considered can be seen as the (bosonic part of the) low-energy effective action of ${\cal N}=2$ supersymmetric $SU(N)$, with $N=n+1$ and with  $N_{f}=n$ flavors.    The gauge group is broken by 
 the adjoint scalar
VEV,
\be  \brc  \phi   \ckt = diag ~(m_{1}, m_{2}, \ldots, m_{n}, - \sum_{j=1}^{n} m_{j})\;, \qquad  m_{i} \to m\;, 
\label{semiclvac}  \ee
to $SU(n) \times U(1)/{\mathbbm Z}_{n} \sim U(n).$
The light monopoles and the magnetic gauge quantum numbers of these, in the limit of small $m_{i}$ and $\mu$,  can be read off from the singularities of the Seiberg-Witten curves \cite{APS,CKM}.  Semi-classically (large $m_{i}$),  instead,  the vacua of this theory are classified according to the number of quark flavors which remain massless due to the cancellation between the bare quark mass and the adjoint scalar VEV
in the superpotential,
\[     \, {\tilde Q}  \, (\sqrt{2}  \, \Phi  + M ) \, Q\;.
\]
The model considered  in \cite{ABEKY,SY,HT2}, as can be seen from the VEV of the adjoint scalar,  corresponds to the $r=n=N_{f}$ vacuum of the above theory.   The light monopoles in Table~\ref{tabsun}  correspond to the limit $m_{i} \to m \to 0$, and we need to know 
to which quantum vacuum each semi-classical vacuum corresponds. 
  This problem of matching the semi-classical and fully quantum mechanical  vacua one by one,  has been solved by using the vacuum counting
and by symmetry considerations.  The classical $r$ vacua, $r=0,1,\ldots, N_{f}$ found in the semiclassical regime $|m_{i}| \gg |\mu| \gg \Lambda $ are found to 
correspond \cite{CKM,BKM,MKY} to the quantum $r$ vacua, $r=0,1,\ldots, N_{f}/2$,  as  
\be     \{ r, \,\, N_{f}- r \} \Longleftrightarrow   r, \;      \qquad  r=0,1,\ldots   \le N_{f}/2\;. 
\label{correspond}\ee
where the left hand side stands for the classical vacuum classification. 
Note that the quantum $r$ vacua (with $SU(r)$ non-Abelian magnetic gauge symmetry) exist only up to $r \le N_{f}/2$
for dynamical reasons \cite{Konishi}. Therefore the model considered in \cite{ABEKY,SY,HT2} must correspond to the $r=0$ quantum vacuum. The latter is characterized by the fact that all monopoles are Abelian (see Table~\ref{tabsun}); furthermore none of them carries any flavor $SU(N_{f})$ quantum numbers.  The condensation of the light monopoles (which occurs when the adjoint scalar masses $\mu \,\Phi^{2}$ are added in the theory)  does not break $SU(N_{f})$ symmetry, consistently with the finding from the vortex dynamics.\footnote{The authors thank R. Auzzi and G. Marmorini for discussions on this point.  }

On the other hand, one knows \cite{APS,CKM} that in four dimensional ${\cal N}=2$ supersymmetric QCD there appear light monopoloes carrying non-Abelian charges ($r$ vacua with $2 \le  r  \le N_{f}/2$ in Table~{\ref{tabsun}),  and one wonders whether such truly non-Abelian vortices which do not Abelianize dynamically can be found in some appropriate regime, through which one can identify a  semi-classical origin of the non-Abelian monopoles and the associated vortices.  

We shall show below that such a system can indeed be found.   The underlying model is the same as the one discussed in \cite{ABEKY,CKM}:  an
 ${\cal N}=2$ supersymmetric $SU(N)$ gauge theory   with  $N_{f}= N$ flavors. But the gauge group is broken partially down to  
 $SU(n) \times SU(r) \times U(1)$ gauge symmetry ($N=n+r)$  by the adjoint scalar VEV.  

  \begin{table}[h]
  {\small 
\begin{center}
\vskip .3cm
\begin{tabular}{ccccc}
  $r$      &   Deg. Freed.      &  Eff. Gauge  Group
&   Phase    &   Global Symmetry     \\
\hline
$0$      &    monopoles   &   $U(1)^{N-1} $               &   Confinement
   &      $U(n_f) $            \\ \hline
$ 1$          & monopoles         & $U(1)^{N-1} $        &
Confinement       &     $U(N_f-1) \times U(1) $        \\ \hline
$ 2,.., [{N_f -1\over  2}] $   &  NA monopoles        &    $SU(r)
\times U(1)^{N-r}   $  &    Confinement
&          $U(N_f-r) \times U(r) $
\\ \hline
$ {N_f / 2}  $    &   rel.  nonloc.     &    -    &    Almost SCFT
&          $U({N_f / 2} ) \times U({N_f/2}) $
\\ \hline
\end{tabular}
\caption{ \footnotesize Confining vacua of $SU(N)$ gauge theory with $N_f$ flavors.   In the superconformal  $r= N_{f}/2$ vacuum,  relatively nonlocal monopoles and dyons appear both as the low-energy effective
degrees of freedom. ``Almost SCFT" means
that the theory is a non-trivial superconformal theory when $\mu=0$ but
confines upon $\mu \neq 0$ perturbation.  In the theory with $N_{f}=N$ considered here,  the vacua at the ``baryonic root'',  in free magnetic phase,  are absent.  They appear only for $N_{f}> N$, with an effective gauge group, $SU(N_{f}-N)$. 
}
\label{tabsun}
\end{center}
}
\end{table}

\section{Dynamical Abelianization} 

As the question of dynamical Abelianization is central to this work, and as this point might be somewhat misleading,  
 let us add a few clarifying remarks before proceeding,  even risking the vice of over-repetition.\footnote{We thank the referee of the first version of this paper for urging us to do so.}     Dynamical Abelianzation, as normally understood, concerns the {\it gauge} symmetry. 
It means by definition that a non-Abelian gauge symmetry of a given theory     reduces at low energies by quantum effects to an Abelian (dual or not) gauge theory.  (Related concepts are dynamical Higgs mechanism, or tumbling \cite{Tumbling}).   
Example of the theories in which this is known to occur are the pure ${\cal N}= 2$ supersymmetric Yang-Mills theories \cite{SW1,curves}  which reduce to Abelian gauge theories at low energies, and  the  $SU(2)$
 ${\cal N}= 2$  theories with $N_{f}=1,2,3$ matter hypermultiplets \cite{SW2}. 
 But as has been emphasized repeatedly and in Introduction above, ${\cal N}= 2$  supersymmetric $SU(N)$ QCD (with $N\ge 3$) {\it  with quark multiplets},  do not Abelianize in general \cite{APS,CKM,Konishi}.   Whether or not the standard QCD  with light quarks Abelianizes is not known.  The 't Hooft-Mandelstam scenario implies a sort of dynamical Abelianization,  as it assumes the Abelian $U(1)^{2}$ monopoles to  be the dominant degrees of freedom at some relevant scales, but this has not been proven. 
 
    As the vortex orientation fluctuation modes turn out to be intimately connected to the way {\it dual } gauge symmetry emerges at low-energies  (\cite{Eto:2006dx,Konishi} and below),   it is perfectly reasonable to use the same terminology for the vortex modes.
 
    Nevertheless, one could  {\it define} -- and  in this paper  we shall use it  in this  sense --   the concept of {\it non-Abelian or Abelian vortices}, independently of the usual meaning attributed to it in relation to a gauge symmetry.   A vortex is non-Abelian, if it carries a non-trivial, internal  non-Abelian moduli, which can fluctuate along its length and in time.  We exclude from this consideration other vortex moduli   associated with their (transverse) positions, shapes  or sizes (in the case of higher-winding \cite{HT2,ASY,Eto:2006cx}  or semi-local vortices \cite{Tong:2005un,Shifman:2006kd,SemiSeven}).     Otherwise, a vortex is Abelian.  The standard ANO vortex is Abelian, as it possesses  no-continuous moduli.  The vortices found in the context of $U(N)$  models \cite{HT,ABEKY}   {\it are } indeed non-Abelian in this sense.

    But just as a non-Abelian gauge theory may or may not Abelianize depending on  dynamical details, a non-Abelian vortex may or may not dynamically Abelianize.  In the very papers in which these vortices have been discovered \cite{ABEKY,HT2}  and in those which followed \cite{SY},  it was shown that they  dynamically reduced to Abelian, ANO like vortices at long distances. The orientational moduli  fluctuate strongly and at long distances they effectively lose their orientation.  A recent observation \cite{SYLues}  nicely  exhibits this aspect through the L\"uscher term of the string tension.  It is quite sensible therefore to call  those vortices in the $U(N)$, $N_{f}=N$ models as {\it elementary} non-Abelian vortices \cite{SYbook}.

In what follows, it will be shown  that this fate is not unavoidable.   Semi-classical 
non-Abelian vortices which remain so at low-energies do exist;  they can be found in appropariate vacua, selected by a careful tuning of the bare quark masses. This is quite similar to the situation in ${\cal N}=1$ supersymmetric QCD, where a vacuum with a prescribed chiral  symmetry breaking pattern can be selected out of the degenerate set of vacua by appropriately tuning the bare quark mass ratios, before sending them to zero. The symmetry breaking pattern in those theories is aligned with the bare quark masses, as is well-known \cite{Kanomaly}.

And this finding closes the gap in matching the results in the $4D$ theories at fully quantum regimes (where all bare mass parameters are small) and in semi-classical regimes where the vortices can be reliably studied.  In other words the work which follows allows us to identify the semi-classical origin of the quantum non-Abelian monopoles found in \cite{APS,CKM}.

\section{Non-Abelian vortices which do not dynamically reduce to ANO vortices } 

The model on which we shall base our consideration  is  the softly broken ${\cal N}=2$  supersymmetric QCD with $SU(N)$ and $N_{f}=N$ flavors of quark multiplets,  
\be
{\cal L}=     \frac{1}{ 8 \pi} {\im} \, {\tau}_{cl} \left[\int d^4 \theta \,
\Tr\,(\Phi^{\dagger} e^V \Phi e^{-V}) 
+\int d^2 \theta\,\frac {1}{ 2}\Tr\,(W W)\right]
+ {\cal L}^{({\rm quarks})}  +  \int \, d^2 \theta \,\mu  \,\Tr \,  \Phi^2;
\label{lagrangian}
\eeq
\beq {\cal L}^{({\rm quarks})}= \sum_i \, \left[ \int d^4 \theta \,  ( Q_i^{\dagger} \, e^V \, 
Q_i + {\tilde Q_i} \,  e^{-V}\,  {\tilde Q}_i^{\dagger} ) +  \int d^2 \theta
\,  ( \sqrt{2}\,  {\tilde Q}_i \, \Phi \, Q^i    +      m_{i} \,    {\tilde Q}_i \, Q^i   )  \right]\;.
\ee
where  $\tau_{cl} \equiv  {\theta_0 / \pi} + {8 \pi i }/{ g_0^2}$ contains the coupling constant and the theta parameter,  $\mu$ is the adjoint scalar mass,  breaking softly  ${\cal N}=2$ supersymmetry to ${\cal N}=1$. 
We tune the bare quark masses as 
\[  m_{1}= \ldots = m_{n} = m^{(1)}; \quad m_{n+1}= m_{n+2}= \ldots = m_{n+r}= m^{(2)}\, , \qquad N=n+r\;;  
 \] 
\be    n\, m^{(1)} +  r\, m^{(2)} =0\;, \label{masses} \ee
or
\be   m^{(1)}=  \frac{r\, m_{0}}{\sqrt{ r^{2} + n^{2}}}, \quad   m^{(2)}= - \frac{ n \, m_{0}}{\sqrt{ r^{2} + n^{2} }},
\label{masscond}\ee
and their magnitude is taken as 
 \be   |m_{0}| \gg  |\mu|  \gg  \Lambda\,.\label{masssemi}
\ee
  The adjoint scalar VEV can be taken to be 
\be \brc \Phi \ckt   =  - \frac{1}{\sqrt{2}}  \left(\begin{array}{cc}m^{(1)} \, {\mathbbm 1}_{n \times n} & {\bf 0} \\ {\bf 0}  & m^{(2)} \, {\mathbbm 1}_{r \times r}  \end{array}\right)  \label{ourvac}
\ee
Below the mass scale  $v_{1} \sim   |m_{i}|$   the system thus  reduces to a gauge theory with gauge group 
\be   G =   \frac {SU(n) \times SU(r) \times U(1)}{{\mathbbm Z}_{K}}\,,   \quad K = {\rm LCM} \, \{n,r\}\;    \label{Gaugegr} 
\ee
 where $K$ is the least common multiple of $n$ and $r$.   The higher $n$ color components of the first $n$ flavors  (with the bare mass $m^{(1)}$)   remain massless, as well as the lower $r$ color components of the last $r$ flavors
 (with the bare mass $m^{(2)}$):  they will be denoted as $q^{(1)}$ and $q^{(2)},$
 respectively.   They carry  the charges $\lambda_{1}, - \lambda_{2}$, 
 \be \lambda_{1} \equiv   \frac{r}{\sqrt{2 \,n\, r\,  ( r + n ) }}; \qquad 
\lambda_{2} \equiv   \frac{n}{\sqrt{2 \,n\, r\,  ( r + n ) }}\;.
\ee
 with respect to the   $U(1)$ gauge symmetry  generated by 
\be   t^{(0)} =    \left(\begin{array}{cc}   \lambda_{1}  \, {\mathbbm 1}_{n \times n} & {\bf 0} \\ {\bf 0} & -  \lambda_{2}   \, {\mathbbm 1}_{r \times r}  \end{array}\right)\;, \qquad  \Tr\,  t^{(0)\, 2} =\frac{1}{2}\;.\label{U1gene} \ee
Non-Abelian gauge groups are generated by the standard $SU$ generators 
\be    t_{su(n)}^{a}=    \left(\begin{array}{cc}  (t^{a})_{n \times n} & {\bf 0} \\ {\bf 0} &  {\bf 0}_{r \times r}  \end{array}\right)\;; \qquad  t_{su(r)}^{b}=    \left(\begin{array}{cc} {\bf 0}_{n \times n}  & {\bf 0} \\ {\bf 0} &  (t^{b})_{r \times r}  \end{array}\right)\;; 
\ee
$a=1,2,\ldots, n^{2}-1$;  $b=1,2,\ldots, r^{2}-1$, with the normalization 
\[  \Tr_{n}  \,( t^{a} \, t^{a^{\prime}} )  = \frac{ \delta_{a a^{\prime}}}{2} , \quad   \Tr_{r}  \, (t^{b} \, t^{b^{\prime}}) = \frac{ \delta_{b b^{\prime}}}{2} \;.  
\]
Our model for studying the vortices  then is:  \footnote{One could very well start with a model of this sort directly.   The squark VEVs can be induced by a Fayet-Iliopoulos term introduced by hand.   By an $SU_{R}(2)$ rotation, which rotates  $(q, {\tilde q}^{\dagger})$ as a doublet, such a model can be seen to be equivalent to the one being considered here. }
\bqa  && {\cal L} =  - \frac{ 1 }{4 g_{0}^2}  F_{\mu \nu}^{0\, 2}  - \frac{ 1 }{4 g_{n}^2}  F_{\mu \nu}^{n\, 2} - \frac{ 1 }{4 g_{r}^2}  F_{\mu \nu}^{r\, 2}   +    \frac { 1}{ g_{0}^2}  |{\cal D}_{\mu} \Phi^{(0)}|^2 +  \frac { 1}{ g_{n}^2}  |{\cal D}_{\mu} \Phi^{(n)}|^2 +  \frac { 1}{ g_{r}^2}  |{\cal D}_{\mu} \Phi^{(r)}|^2 \non \\
&& +     \left|{\cal D}_{\mu}  q^{(1)} \right|^2 + \left|{\cal D}_{\mu} \bar{\tilde{q}}^{(1)}\right|^2  + 
  \left|{\cal D}_{\mu}  q^{(2)} \right|^2 + \left|{\cal D}_{\mu} \bar{\tilde{q}}^{(2)}\right|^2     -    V_D -  V_F, 
\label{Lag}\eea
plus fermionic terms,  
where  $V_{D}$ and $V_{F}$ are the $D$-term and $F$-term potentials.    The $D-$term potential $V_{D}$ has the form, 
 \be V_{D}   =          \frac { 1}{ 8 }  \, \sum_A  \left(\Tr\, t^A \,[
 \,  \frac { 2}{ g^2 }  \, [\Phi,   \Phi^\dagger]  +  
\sum_{i}(Q_i   Q_i^\dagger -  {\tilde Q}_i^\dagger   {\tilde Q}_i )\,] \right)^2;
\ee
where  the generators $A$ takes the values  $0$ for $U(1)$,  $a=1,2, \ldots, n^{2}-1$ for $SU(n)$ and  $b=1,2, \ldots r^{2}-1$ for $SU(r)$.
 $V_F$ is of  the form 
\bqa  &&   g_{0}^2 \, |   \mu \, \Phi^{(0)} +
\sqrt 2   \, {\tilde Q} \, t^{(0)} \, Q |^2    +    g_{n}^2 \, |   \mu \, \Phi^{(a)} +
\sqrt 2   \, {\tilde Q} \, t_{su(n)}^{(a)} \, Q |^2    +   g_{r}^2\, |   \mu \, \Phi^{(b)} +
\sqrt 2   \, {\tilde Q} \, t_{su(r)}^{(b)} \, Q |^2  \non \\
&&  +     {\tilde Q } \,   [ \, M    + \sqrt2  \Phi   ] \,  [ \, M    + \sqrt2 \, \Phi   ]^{\dagger}  \, {\tilde Q}^{\dagger} +   Q^{\dagger} \, [ \, M    + \sqrt2  \, \Phi   ]^{\dagger}    \, [ \, M    + \sqrt2 \,  \Phi   ]  \, Q\,, 
\label{VFpotential}\eea
where 
\[    M=  \left(\begin{array}{cc}m^{(1)} \, {\mathbbm 1}_{n \times n} & {\bf 0} \\ {\bf 0}  & m^{(2)} \, {\mathbbm 1}_{r \times r}  \end{array}\right)
\]
is  the mass matrix and the (massless) squark fields have the form, 
\be   Q(x) =   \left(\begin{array}{c c}  q^{(1)}(x) _{n\times n}  & 0 \\ 0  &  q^{(2)}(x)_{r \times r}\end{array}\right)\,,\quad   {\tilde Q}(x) =   \left(\begin{array}{c c}  {\tilde q}^{(1)}(x) _{n\times n}  & 0 \\ 0  &  {\tilde q}^{(2)}(x)_{r \times r}\end{array}\right)\,, \ee
if written in a color-flavor mixed matrix notation.   The light squarks (supersymmetric partners of the left-handed quarks in supersymmetric model)  are summarized in Table~\ref{lightq}. 
\begin{table}
\begin{center} 
  \begin{tabular}{| c c c c | }
\hline
 fields  &    $U(1)$ & $SU(n)$  & $SU(r)$    \\
 \hline 
  $q^{(1)}$   &  $\lambda_{1}$ &   ${\underline n}$ &  ${\underline 1}$\\
  ${\tilde q}^{(1)} $ &  $- \lambda_{1}$ &  $ {\underline n^{*}} $&  ${\underline 1}$ \\
   $q^{(2)}$   &$  - \lambda_{2}$& ${\underline 1} $ &  ${\underline r} $ \\
  ${\tilde q}^{(2)}$  & $\lambda_{2}$  & ${\underline 1}$  & ${\underline r^{*}}$   \\
\hline
\end{tabular}
\end{center}
  \caption{ }\label{lightq}
\end{table}

We set $V_{D}$   to zero identically, in the vacuum and in the vortex configurations, by  keeping 
\be   {\tilde{q}}^{(1)}  =  (q^{(1)})^{\dagger}, \qquad  {q}^{(2)}  =  -  ({\tilde q}^{(2)})^{\dagger}\;;
\label{reduction1}   \ee
the redefinition   
\be  q^{(1)} \to \frac{1}{\sqrt{2}} \, q^{(1)}, \quad  {\tilde q}^{(2)} \to \frac{1}{\sqrt{2}} \, {\tilde q}^{(2)}
\label{reduction2}   \ee
 brings the kinetic terms for these fields back to the original  form. 

The VEVs of the adjoint scalars are  given by 
\be   \brc \Phi^{(0)} \ckt =-  m_{0}, \qquad \brc \Phi^{(a)} \ckt = \brc \Phi^{(b)} \ckt   =0,\; \label{adjVev}
\ee
while the squark VEVs are given (from the vanishing of the first line of Eq.~(\ref{VFpotential}))    by
\be    
\brc Q \ckt =   \left(\begin{array}{c c}  v^{(1)} \, {\mathbbm 1}_{n\times n}  & 0 \\ 0  &  - v^{(2)\, *} \, {\mathbbm 1}_{r\times r}  \end{array}\right)\,,\quad \brc {\tilde Q} \ckt =   \left(\begin{array}{c c}  v^{(1)\, *} \, {\mathbbm 1}_{n\times n}  & 0 \\ 0  &     v^{(2)} \, {\mathbbm 1}_{r\times r}  \end{array}\right)\,,
\label{QkVEV}\ee
with
\be 
 | v^{(1)}|^{2} +    | v^{(2)}|^{2}  = \sqrt{\frac {n+r}{n\, r}} \,  \mu \, m_{0}\,  \;. 
\ee
There is a continuous vacuum degeneracy;  we assume that 
\[    v^{(1)} \ne 0; \qquad  v^{(2)} \ne 0\;,
\]
in the following.  The presence of the flat direction implies the existence of the so-called semi-local  vortex moduli; but we shall not  be concerned with these here. 

``Non-Abelian'' vortices exist in this theory as the vacuum breaks the gauge group $G$ (Eq.~(\ref{Gaugegr})) completely, leaving at the same time a color-flavor diagonal symmetry 
\be   [ SU(n) \times SU(r) \times U(1) ]_{C+F}  
\label{colorflavor}\ee
 unbroken.   The full global symmetry, including the overall global $U(1)$ is given by 
 \be   U(n) \times U(r)\;.  \label{notethat}
 \ee
   The minimal vortex in this system corresponds to the smallest nontrivial loop in the $G$ group space,  Eq.~(\ref{Gaugegr}).
It is the path in the $U(1)$ space
\be   \left(\begin{array}{cc} e^{i \alpha r}  {\mathbbm 1}_{n \times n} & 0 \\0 & e^{i \alpha n} {\mathbbm 1}_{r \times r }\end{array}\right)\,,\quad \alpha  : 0 \to    \frac{2\pi }{ n\, r },
\ee
that is, 
\be  {\mathbbm 1}_{N \times N} \to    {\mathbbm Y}, 
\qquad      {\mathbbm Y} = \left(\begin{array}{cc}e^{2 \pi i/n } {\mathbbm 1}_{n \times n} & 0 \\0 & e^{ 2 \pi i / r} {\mathbbm 1}_{r \times r }\end{array}\right)\,, 
\ee
followed by a path 
 in the $SU(n) \times SU(r)$ manifold 
 \be   {\mathbbm 1}_{n\times n} \to  {\mathbbm Z}_{n}=  e^{
-\frac{2 \pi i}{n} } {\mathbbm 1}_{n\times n} ; \qquad 
{\mathbbm 1}_{r\times r} \to   {\mathbbm Z}_{r}= e^{-\frac{2 \pi i}{r} } {\mathbbm 1}_{r\times r} ;\ee
 back to the unit element. For instance one may choose 
  ($\beta:0 \to 2\pi; \gamma:0\to 2\pi$)
 \[  \left(\begin{array}{cc}e^{i\beta (n-1) /n} & 0 \\ 0 & e^{-i \beta /n}\, {\mathbbm 1}_{(n-1)\times (n-1)}\end{array}\right)\;;\quad 
 \left(\begin{array}{cc}e^{i\gamma (r-1) /r} & 0 \\ 0 & e^{-i \gamma /r}\, {\mathbbm 1}_{(r-1)\times (r-1)}\end{array}\right)\;.
  \]
   As   
 \be  {\mathbbm Y}^{K}  =   {\mathbbm 1}_{N \times N},  \quad K = {\rm LCM} \, \{n,r\}\;
 \ee
 it follows that the tension (and the winding) with respect to the $U(1)$ is   $\frac{1}{K} $ of that in the standard ANO  vortex.  

The squark fields trace such a path asymptotically, i.e., far  from the vortex core,
as one goes around the vortex;    at finite radius the vortex has, for instance, the form,  
\be   q^{(1)}  =   \left(\begin{array}{c c    }  e^{i \phi} \,f_{1}(\rho)  &  0   \\
0 &   f_{2}(\rho)\, {\mathbbm 1}_{(n-1)\times (n-1)}  
 \end{array}\right)\,, \quad 
 {\tilde q}^{(2)}  =   \left(\begin{array}{c c    }  e^{i \phi} \,g_{1}(\rho)  &  0   \\
0 &  g_{2}(\rho)\, {\mathbbm 1}_{(r-1)\times (r-1)}  
 \end{array}\right)\,, 
\label{vortexconf} \ee
 where $\rho$ and $\phi$ stand for the polar coordinates in the plane perpendicular to the vortex axis,  $f_{1,2}, g_{1,2}$ are profile functions.  The adjoint scalar fields $\Phi$ are taken to be equal to their VEVs, Eq.~(\ref{adjVev}).  They are accompanied by the appropriate gauge fields so that the  tension is finite.   The BPS equations for the squark and gauge fields, and the properties of their solutions are discussed in Appendix A.  The behavior of numerically integrated vortex profile functions $f_{1,2}, g_{1,2}$ is illustrated in Fig.~\ref{Vortexprofile}. 
 \begin{figure}
\begin{center}
\includegraphics[width=2.5 in]{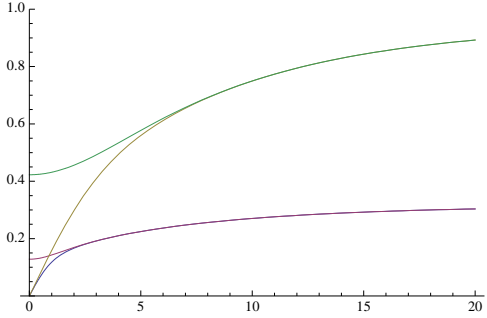}
\caption{\footnotesize Numerical result for the profile functions $f_{1,2}, g_{1,2}$ as functions of the radius $\rho$, for $SU(3)\times SU(2)\times U(1)$  theory.  The coupling constants and the ratio of the VEVs are taken to be $g_{0}=0.1$, $g_{3}=10$, $g_{2}=1$, $|v_{2}|/ |v_{1}|=3$.  }
\label{Vortexprofile}
\end{center}
\end{figure}

   We note here only that the necessary boundary conditions on the squark profile functions have the form, 
   \[     f_{1}(\infty)=  f_{2}(\infty) = v^{(1)}, \qquad  g_{1}(\infty)= g_{2}(\infty) = v^{(2)}, 
   \]
   while at the vortex core, 
   \be   f_{1}(0)=0, \quad  g_{1}(0) =0, \qquad   f_{2}(0) \ne 0, \quad  g_{2}(0) \ne 0,
  \label{fundamental} \ee
  
  The most important fact about these minimum vortices  is that  one of the $q^{(1)}$ {\it and}  one of the ${\tilde q}^{(2)}$ fields must necessarily wind at infinity, simultaneously.  
   As the individual vortex breaks the (global) symmetry of the vacuum as 
\be  [ SU(n) \times SU(r) \times U(1) ]_{C+F}  \to  SU(n-1)  \times SU(r-1)  \times U(1)^{3}, 
\label{smaller}  \ee
the vortex acquires  Nambu-Goldstone modes parametrizing
\be   CP^{n-1} \times CP^{r-1}\;:         
\ee
they transform under the exact color-flavor symmetry $SU(n) \times SU(r)$  as the 
bi-fundamental  representation, $({\underline n}, {\underline r})$.
Allowing the vortex orientation to fluctuate along the vortex length and in time, we get a $CP^{n-1} \times CP^{r-1}$  two-dimensional sigma model as an effective Lagrangian describing them.  The details have been worked out in \cite{SY,HT2}  and  need not be repeated here.

The main  idea of the present paper is this.  Let us assume without losing generality that  $n > r$, excluding the special  case of $r=n$  for the moment.  As has been shown in \cite{SY,HT2} the coupling constant of the $CP^{n-1}$  sigma models grows precisely as the  coupling constant of the $4D$ $SU(n)$ gauge theory.   At the point the $CP^{n-1}$  vortex moduli fluctuations become strong and 
the dynamical scale $\Lambda$ gets generated, with vortex kinks (Abelian monopoles) acquiring mass of the order of  $\Lambda$, 
the vortex still carries the unbroken $SU(r)$  fluctuation modes ($CP^{r-1}$),  as 
the $SU(r)$ interactions are still weak.  See Fig.~\ref{sunsur}.
\begin{figure}
\begin{center}
\includegraphics[width=3in]{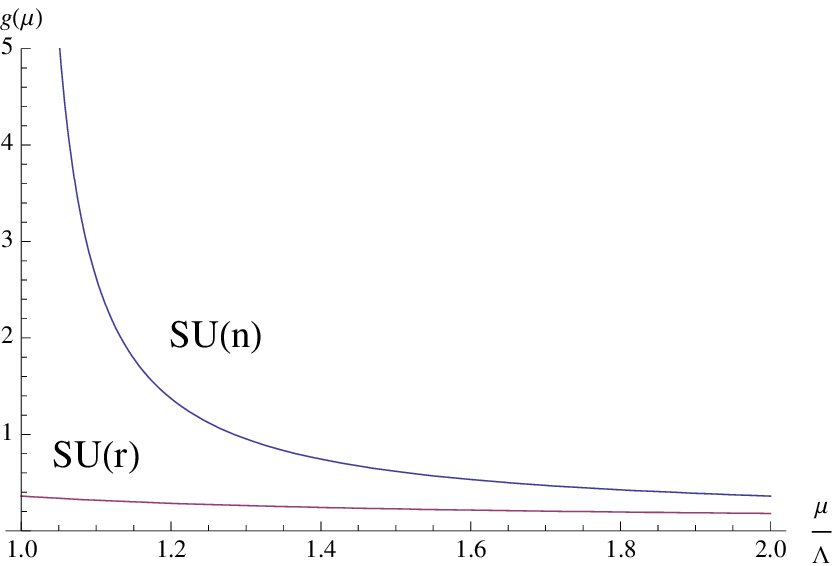}
\caption{ }
\label{sunsur}
\end{center}
\end{figure}
  Such a vortex will carry one of the $U(1)$ flux 
arising from the dynamical breaking of $SU(n) \times U(1) \to  U(1)^{n}$, as well as an $SU(r)$ flux.  
As these vortices end at a massive monopole (arising from the high-energy gauge symmetry breaking, Eq.~(\ref{ourvac})),  the latter necessarily carries a non-Abelian continuous moduli, whose points transform as in
the fundamental representation of 
$SU(r)$.  This can be  interpreted as the  (electric description of)  dual gauge $SU(r)$ system observed in the infrared limit of the $4D$ SQCD \cite{APS,CKM}.  

The special case $r=1$ corresponds to the $U(N)$ model  \cite{ABEKY,SY,HT2,Eto:2006pg}, mentioned in the Introduction, and in this case the vortices dynamically Abelianize.  This is not in contradiction with the claim made above, after Eq.~(\ref{correspond}), that the $U(n)$ models considered in those papers corresponded to the quantum $r=0$ vacuum  of the $SU(n+1)$ model, with  $N_{f}=n$.  The point is that here we start with the underlying theory with $SU(N)$, $N_{f}=N$, where $N=n+r$;  the classical-quantum vacuum matching condition  (Eq~.(\ref{correspond})) implies that the $U(n)$ models studied earlier, if embedded in our general scheme,  correspond to the $r=1$, rather than $r=0$, vacua. The symmetry breaking pattern 
 Eq.~(\ref{notethat}) also perfectly matches the full quantum result in Table~\ref{tabsun}, as it does for generic $r$.

There is no difficulty in generalizing our construction and finding  vortices with fluctuations 
corresponding to more than two non-Abelian factors, 
\[   SU(n) \times SU(r_{1}) \times SU(r_{2}) \times \ldots \;, \]
as long as we remain in the semi-classical region with $|m_{i}|, |\mu| \gg  \Lambda. $  However, the main aim of this paper is to identify the semi-classical origin of the non-Abelian monopoles 
seen in the fully quantum effective low-energy action of the theory at $m_{i} \to 0$, $\mu \sim \Lambda$.    
In such a limit,  the breaking of the gauge symmetry is a dynamical question;    the result of the analysis of the $4D$ theory (Table~\ref{tabsun}) suggests that  in that limit the surviving non-Abelian dual group $SU(r_{1}) \times SU(r_{2}) \times \ldots$ gets enhanced to a single factor $ SU(r)$.  In order for gauge groups with more than one non-Abelian factors to survive dynamically, a nontrivial potential in the adjoint scalar field $\Phi$ needs to be present in the underlying theory \cite{BKM}.

\section{Vortex moduli, kinks and monopoles in $4D$ theory} 

It is somewhat a puzzle why the exact $2D$-$4D$ correspondence holds. A particularly intriguing point is that the two-dimensional vortex sigma-model dynamics in the {\it Higgs phase} of the four dimensional theory  reproduces exactly the $4D$ gauge dynamics in the {\it Coulomb phase}.  One might be tempted to argue that the reason for such a correspondence is that in the vortex core the full gauge symmetry is restored, as in the case of an instanton.  Actually, it is not.
A glance at Eq.~(\ref{vortexconf}) and Eq.~(\ref{fundamental})  shows that the {\it gauge} symmetry at the vortex core is only partially restored, to  $U(1) \times U(1)$.    The {\it global} symmetry in the vortex core, on the other hand,  is smaller than that outside the vortex
(Eq.~(\ref{smaller})).  This difference in the global symmetries means that there are certain Nambu-Goldstone excitations (and their superpartners) which can propagate only inside the vortex.  In the vacuum exterior to the vortex these modes become massive and cannot propagate.   They correspond to the various broken $SU(n)_{C+F}\times SU(r)_{C+F}$ generators, 
\be \left(\begin{array}{c c c c    } 0  &  {\mathbbm  b}^{\dagger}   & 0 & 0 \\
 {\mathbbm  b}   &    {\mathbf 0}_{(n-1)\times (n-1)}  & 0 &  0   \\ 
0   & 0  &   0   & {\mathbbm  c}^{\dagger} \\
0 & 0 &  {\mathbbm c}  &     {\mathbf 0}_{(r-1) \times (r-1)}  \end{array}\right)\,,
\ee
with complex $n-1$ component vector $ {\mathbbm  b} $ and $r-1$ component vector $ {\mathbbm  c}.$  They are precisely   the inhomogeneous coordinates of $CP^{n-1}$ and $CP^{r-1}$, respectively, which are the non-Abelian vortex
flux orientation moduli.\footnote{In the strictly low-energy approximation, Eq.~(\ref{Lag}),  where small terms arising from 
the symmetry breaking at high energies are neglected,  the vortices are BPS saturated:  their moduli space turns out to be considerably larger and shows a richer structure. Here we restrict ourselves to the vortex moduli arising form the global symmetry alone.  The latter is an exact symmetry of the system, valid in the full theory, while most of the moduli in the BPS approximation will be absent in the exact theory.  As  emphasized in \cite{Strassler,Eto:2006dx} the fact that the high-energy monopoles and low-energy vortices are  both approximately BPS {\it but not exactly so}, is fundamental in the monopole-vortex matching argument. 
}

   In our opinion, the true reason for the exact $2D$-$4D$  correspondence is in the consistency of being able to consider the model for the vortex, such as  Eq.~(\ref{Lag})  or similar models with $U(n)$ gauge symmetry,  as a low-energy approximation of (e.g.)  an  $SU(N)$ gauge theory, $N> n$.   
The fact  that $\Pi_{2}(SU(N)) = {\mathbbm 1}$  means that any regular 't Hooft-Polyakov monopoles arising from a partial breaking such as $SU(N) \to SU(n) \times SU(r)\times U(1)$ at an intermediate mass scale, must eventually all disappear from the spectrum, confined by the vortices developing at the lower energies, when much smaller squark VEVs are taken into account.  Vice versa, no vortex appearing in the low-energy $ SU(n) \times SU(r)\times U(1)$ theory in Higgs phase {\it can be there}  in the underlying $SU(N)$  theory: they are meta-stable and must end at the 
massive monopoles. Consistency requires that the vortex with {\it each} orientation must have its counterpart  --  a monopole with the corresponding orientation.   Symmetry-based vortex moduli space implies an associated monopole moduli space.  
 
   Note that the color-flavor diagonal symmetry $U(n)\times U(r)$ is an exact symmetry of the full system (Eq.~(\ref{QkVEV})).  When the low-energy vortex orientation is rotated in the quotient space $CP^{n-1} \times CP^{r-1}$, a corresponding rotation must be performed on the monopole at the extrema, to keep the energy of the configuration invariant.  Although the origin of such  fluctuation modes is color-flavor {\it global} symmetry,  the vortex can end (or originate)  anywhere and at any instant of time into (from) a monopole. 
   (Fig.~\ref{anywhere}).  This could be the reason why these fluctuation modes,  dynamically broken or not,  manifest themselves as a dual {\it local} gauge group.  The latter is realized however in a confining phase, as the original, electric gauge group is in Higgs phase.\footnote{Basically, this is not very different from what happens in the two-dimensional Ising model at the critical temperature, although there the kinks in the spin chain manifest themselves as  massless unconfined fermionic  particles in the dual  picture.  (See for instance, Kogut\cite{Kogut}) for a review.}  The vortex of the electric theory is the confining string of the dual  theory. 
   
\begin{figure}
\begin{center}
\includegraphics[width=2in]{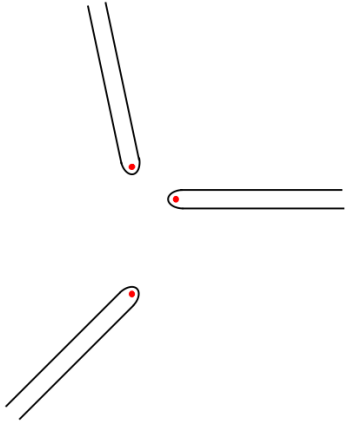}
\caption{ \footnotesize Vortex carrying non-Abelian flux moduli can convert to a monopole anywhere and at any time}
\label{anywhere}
\end{center}
\end{figure}
   
\section{Matching to the $4D$  theory}  

There remains the task of matching the light magnetic degrees of freedom found in the $r$ vacua of the underlying $SU(N)$, $N_{f}=N$  SQCD,  in the  $m_{i} \to 0,$   $\mu  \sim \Lambda$  limit, see Table~\ref{tabnonb},  to the vortices and  their endpoints seen in the low-energy  model (in the region  $|m_{i}|\gg |\mu| \gg \Lambda$).    The vortex carrying $SU(n) \times SU(r) \times U(1)$  quantum numbers, in which $SU(n) $ Abelianizes dynamically to $U(1)^{n-1}$,  so that the monopoles at which these vortices end carry the quantum numbers of $SU(r) \times U(1)$,  is an excellent candidate to explain the   
 appearance of the non-Abelian monopoles in the infrared in $4D$ theory \cite{APS,CKM}.  The fact that both in $4D$ and in $2D$ these solitons exist only for $r 
 \le  N_{f}/2$  is a strong indication that such an identification is indeed correct.

  The fact that the monopoles carrying the $SU(r)$ charge
 appear $N_{f}$ times and represent the global $SU(N_{f})$ symmetry group (see Table~\ref{tabnonb}), is important for the $4D$ low-energy effective action to possess the correct {\it global} symmetry group of the underlying theory \cite{CKM,MKY}.  From the semi-classical point of view, this can be understood as due to the Jackiw-Rebbi effect \cite{JR,BK}. Note that due to this effect, the dual $SU(r)$ group of the fully quantum mechanical regime,  $|m_{i}|, |\mu| \ll  \Lambda$,  is infrared free.\footnote{This is precisely the reason why such monopoles carrying non-Abelian charges can appear as the low-energy degrees of freedom. }

  In the semi-classical region,  $|m_{i}|\gg |\mu| \gg \Lambda$, where we study the vortices,  the Jackiw-Rebbi effect is due to the quark clouds (normalizable fermion zeromodes in the quantization around the background semiclassical monopoles), of the size of    $\sim 1/ |m|  \ll    1/\Lambda. $  We claim that these are effects {\it distinct}  from the color-flavor symmetry breaking effect,
 which involves a much larger length scale of the order of $1/ \sqrt{\mu \,m} $, and  which,  we believe, explains the origin of the dual {\it gauge} group.

\begin{table}[h]
\begin{center}
\vskip .3cm
\begin{tabular}{ccccccc}

&   $SU(r)  $     &     $U(1)_0$    &      $ U(1)_1$
&     $\ldots $      &   $U(1)_{n-1}$    &  $ U(1)_B  $  \\
\hline
$n_f \times  q$     &    ${\underline {\bf r}} $    &     $1$
&     $0$
&      $\ldots$      &     $0$             &    $0$      \\ \hline
$e_1$                 & ${\underline {\bf 1} } $       &    0
&
1      & \ldots             &  $0$                   &  $0$  \\ \hline
$\vdots $  &    $\vdots   $         &   $\vdots   $        &    $\vdots   $
&             $\ddots $     &     $\vdots   $        &     $\vdots   $
\\ \hline
$e_{n-1} $    &  ${\underline {\bf 1}} $    & 0                     & 0
&      $ \ldots  $            & 1                 &  0 \\ \hline
\end{tabular}
\caption{\footnotesize The effective low-energy degrees of freedom and their quantum numbers at the confining vacuum characterized by a magnetic dual $SU(r)$ gauge group.   }
\label{tabnonb}
\end{center}
\end{table}

The Abelian monopoles seen as kinks in the low-energy vortex theory might be identified with the Abelian monopoles in Table~\ref{tabnonb}.  Note that our argument (the monopoles should not be present in the full theory) applies to these monopoles as well.  Though these monopoles {\it are} stable in $2D$  theory, with the vortex extending along a fixed (e.g., in $z$) direction (the first figure in Fig.~\ref{twovor}), {\it they are not stable}  when such a system is embedded in $4D$ theory: they are  confined by a pair of vortices (the second picture of Fig.~\ref{twovor}).   That each of the vortices on both sides of the kink transforms as ${\underline r}$ of $SU(r)_{C+F}$ group  is not in contradiction with the claim that these (kink) monopoles are singlets of $SU(r)_{C+F}$.  Composite vortices transform   as  in a product representation \cite{Eto:2006cx},  which in our case is: 
\[ {\underline r} \otimes {\underline r}^{*} = {\underline 1} \oplus {\underline {r^{2}-1}}\;;  \]
 it can very well be that the lower-tension \footnote{When higher quantum effects are taken into account the two multiplets  
are expected to split. }  double-vortex belongs to the singlet.  
 
It is interesting to consider the case, $r= N_{f}/2$.  In four dimensional ${\cal N}=2$   SQCD  this is a special vacuum, it is a (strongly-coupled) non-trivial superconformal theory. The infrared degrees of freedom include relatively nonlocal monopoles and dyons, and no effective Lagrangian description is available there. Nevertheless, it has been argued \cite{AGK,MKY}  that these represented an interesting type of confining theory (with $\mu \ne 0$) in which confinement is induced by the condensation of {\it monopole composites}, caused by the strong interactions. The symmetry breaking pattern reflects such a mechanism.  
In two dimensional  vortex effective theory, this particular case deserves indeed further study. 
 
 In order to really sew things up, one must answer the following question: do not 
  $CP^{r-1}$ fluctuations also eventually become strongly coupled, generating still another, hierarchically small, mass scale  $\Lambda^{\prime}$, and Abelianize?  
If the  $SU(n) \times SU(r) \times U(1)$ theory were considered in its own right,   without referring to a $4D$ theory, then the answer would be obviously:  yes.   The new scale at which  $SU(r)$ fluctuations become strongly coupled,  $\Lambda^{\prime}$, however, would depend on the coupling constants $g_{r}$  at the ultraviolet cutoff,  which is an arbitrary parameter. 

Actually, as our  $SU(n) \times SU(r) \times U(1)$ theory  is  {\it a low-energy approximation}  of the underlying $4D$ $SU(N)$ theory,
the above argument does not hold.  We assume that our $2D$ system corresponds to the quantum $r$ vacua, with $r< N_{f}/2$.  Such an identification is justified, apart from the fact that the condition  $r< N_{f}/2$ is needed for both of them,  by the unbroken global symmetry $ U(n) \times U(r)$, common to both of the systems.  The vortex carrying a quantum $CP^{r-1}$ modulation, being unstable,  ends at a monopole before the new scale $\Lambda^{\prime}$ is generated by the strong $CP^{r-1}$ interactions.  

 
In the $2D-4D$ matching,  a subtle role is played by the adjoint mass $\mu$. In our vortex study the Fayet-Iliopoulos term  of the low-energy model (see Eq.~(\ref{le})) is given by the mass $\xi \sim \sqrt{\mu \Lambda}$ which should be taken much larger than $ \Lambda$ to analyse the vortices semi-classically. 
 On the other hand, in the fully quantum regime where $4D$ theory is analyzed by use of the Seiberg-Witten solutions  it is necessary to choose   $\mu \ll   \Lambda$ so that the  dual Higgs phenomenon (for  ${\tilde H}=  SU(r)\times U(1)^{N-r}$)    occurring  at the mass scale $\Lambda^{\prime \prime}\sim \sqrt{\mu \Lambda}$  can be reliably studied \cite{CKM} in an effective low-energy action defined at scales lower than $ \Lambda$.   
 It is not known  whether a more quantitative $2D-4D$ matching procedure eventually allows us to identify the two small scales $\Lambda^{\prime}$ (the scale at which $CP^{r-1}$ becomes strongly coupled in the $2D$ theory) and 
 $\Lambda^{\prime \prime} $ (in the $4D$ theory).  The question is rather subtle, as we are really talking about two different kinematical regions, semi-classical ($m_{i}, \mu \gg \Lambda$) and fully quantum ($m_{i}, \mu \sim   \Lambda$),  of the underlying $4D$  theory. 
 
In any case,   both  in $2D$ and $4D$ theories, the $SU(r)$ group disappears at scales lower than  $\Lambda^{\prime}$  or  $\Lambda^{\prime \prime}.$  Of course,  the emergency of a non-Abelian dual group concerns the mass scales higher than these scales  ($\Lambda^{\prime}$ in the $2D$ theory  or $\Lambda^{\prime \prime} $ in the $4D$ theory).  As  
 dual $SU(r)$ gauge interactions correctly describe the monopole interactions at scales higher than $\Lambda^{\prime \prime}$ in $4D$ theory, 
    there must be some range of mass scales at which vortex modulation modes in $CP^{r-1}$ survive,  at mass scale higher than ${\Lambda}^{\prime}$  but much lower than the scale  of gauge symmetry breaking, 
$SU(n) \times SU(r) \times U(1)  \to   SU(r) \times U(1)^{n}.$
 This is indeed what we have found. 
  
\begin{figure}
\begin{center}
\includegraphics[width=4.5 in]{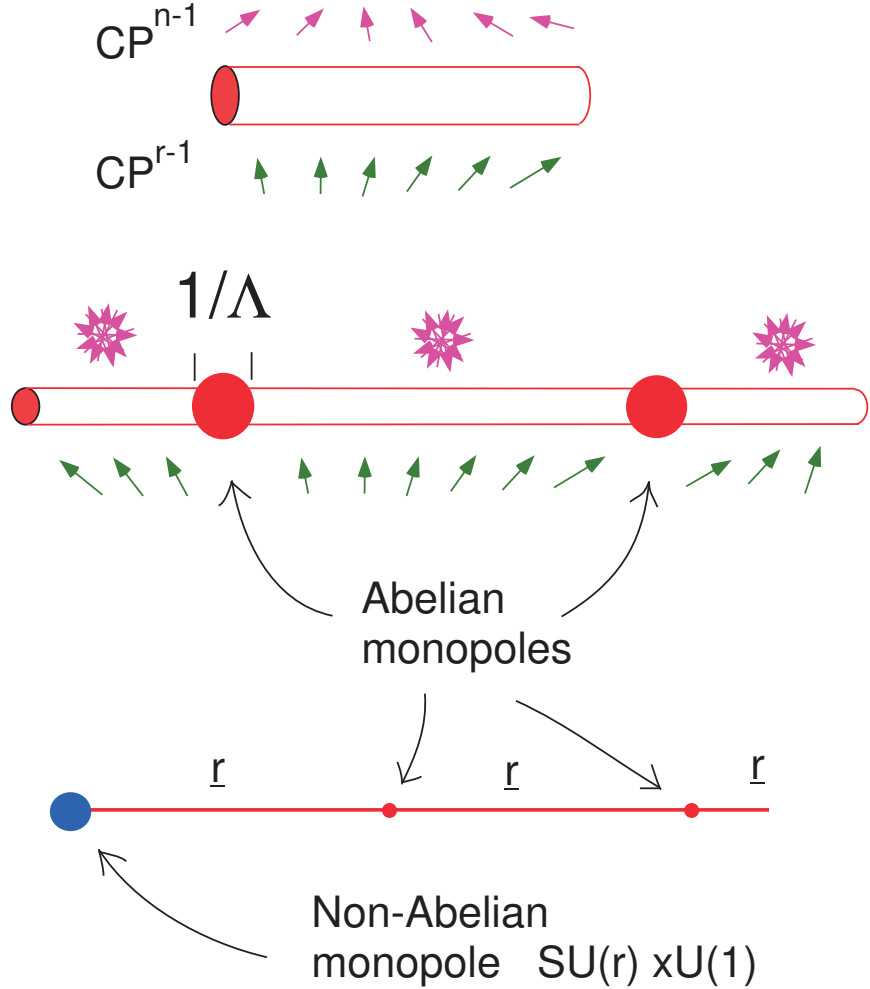}
\caption{\small Our vortex has $CP^{n-1}\times CP^{r-1}$ orientational modes which can fluctuate along the vortex length and in time (top figure).  At low energies $CP^{n-1}$ orientational modes fluctuate strongly and Abelianize, leaving the weakly fluctuating $CP^{r-1}$ modes (middle figure).  The vortex ends at a monopole which, absorbing the $CP^{r-1}$ fluctuations, turns into a non-Abelian monopole.  The latter transforms according to the fundamental representation of the dual $SU(r)$ group (bottom picture).  The kink monopoles are Abelian. }
\label{twovor}
\end{center}
\end{figure} 
  
  \section{Conclusion}
 
In this note we have constructed vortices having non-Abelian moduli, which do not dynamically Abelianize. Semi-classically, they are simply vortices carrying the $SU(n) \times SU(r) \times U(1)$ color-flavor flux.  More precisely,  they carry the Nambu-Goldstone modes 
\[  CP^{n-1}\times CP^{r-1}\;,
\]
resulting from the partial breaking of the $SU(n) \times SU(r) \times U(1)$  global symmetry to $SU(n-1) \times SU(r-1) \times U(1)^{3}$ by the vortex.
For  $n>r$,   $CP^{n-1}$ field fluctuations propagating along the vortex length  become strongly coupled in the infrared, the $SU(n) \times U(1)$ part dynamically Abelianizes;  the vortex however still carries weakly-fluctuating $SU(r)$ flux modulations.  In our  theory  where  $SU(n) \times SU(r) \times U(1)$ model emerges as the low-energy approximation of an underlying $SU(N)$ theory,  such a vortex is not stable. When the vortex ends at a monopole, its  $CP^{r-1}$ orientational modes  are turned into the dual $SU(r)$ color modulations of the monopole.

\section*{Acknowledgement} 

The authors thank Roberto Auzzi, Nick Dorey, Minoru Eto, Sven Bjarke Gudnason, David Tong and Walter Vinci for useful discussions.  This work started when one of us (K.K.) was visiting Isaac Newton Institute for Mathematical Sciences (INI), Cambridge, UK, in the context of the Program, ``Strong Fields, Integrability and Strings''.  K.K. thanks  Nick Dorey and the Workshop organizers for inviting him,  INI for its hospitality and for the stimulating research atmosphere.  
The work of K.O.~is supported by the Research 
Fellowships of the Japan Society for 
the Promotion of Science for Research Abroad.

\appendix

  \section {Vortex configurations}

To be complete we present here the vortex equations and their solutions of the model Eq.~(\ref{Lag}), in the vacuum Eq.~(\ref{adjVev}), Eq.~(\ref{QkVEV}).   
The action of our model, after setting $\Phi$ to its VEV (Eq.~(\ref{adjVev})),  and after making the Ans\"atze-reduction  on the squark fields 
Eqs.~(\ref{reduction1}), (\ref{reduction2}),  takes the form, 
\bqa &&
S =\int d^4x \, \Big [ \frac{1}{4 g_{n}^2} \left(F^{a}_{\mu\nu}\right)^2 + \frac{1}{4 g_{r}^2} \left(F^{b}_{\mu\nu}\right)^2 +
\frac1{4g_{0}^2}\left(F^{(0)}_{\mu\nu}\right)^2
+ \left| {\cal D}_{\mu} q^{(1)} \right|^2  + \left| {\cal D}_{\mu} {\tilde q}^{(2)} \right|^2   \non \\
&&  + \frac{g_{n}^2}{2}\left({q}^{(1)\, \dagger }  t^{a} q^{(1)}  \right)^2+\frac{g_{r}^2}{2}\left({\tilde q}^{(2)} t^{b} {\tilde q}^{(2)\, {\dagger}}   \right)^2+
\frac{g_{0}^2}{2}   \left( \lambda_{1} \, q^{(1) \, \dagger } q^{(1)} +   \lambda_{2} \,  {\tilde q}^{(2)} {\tilde q}^{(2)\, \dagger}    -  \xi
 \right)^2
\Big]\;,  \non \\
\label{le}
\eea
\be 
 \xi = \sqrt{2}\, \mu\, m_{0}\;. 
\ee
The tension can be written completing the squares \`a la
Bogomolny \cite{BPS}, as:
\bea
T &=&\int{d}^2 x \,   \Big (  \,  \sum_{a=1}^{n^{2}-1}  \left[\frac1{2g_n}F^{(a)}_{ij } \pm
     \frac{g_n}{2}
\left({q}^{(1)\, \dagger} t^a q^{(1)}  \right)
\epsilon_{ij} \right]^2+
 \sum_{b=1}^{r^{2}-1}  \left[\frac1{2g_r}F^{(b)}_{ij } \pm
     \frac{g_r}{2}
\left({\tilde q}^{(2)}\,  t^b  \, {\tilde q}^{(2)\, \dagger}  \right)
\epsilon_{ij} \right]^2   \non \\
&+&  \left[\frac1{2g_0}F^{(0)}_{ij} \pm
     \frac{g_0 }{2}
\left( \lambda_{1} \, \Tr_{n}  \, q^{(1)\, \dagger}  q^{(1)}  + \lambda_{2}\,  \Tr_{r}  \, {\tilde q}^{(2)\, \dagger}  {\tilde q}^{(2)}  -  \xi \right)
\epsilon_{ij }\right]^2   \non  \\
&+&  \frac{1}{2} \left|{\cal D}_i \,q^{(1)} \pm   i   \epsilon_{ij}
{\cal D}_j\, q^{(1)}\right|^2   +  \frac{1}{2} \left|{\cal D}_i \,{\tilde q}^{(2)} \pm   i   \epsilon_{ij}
{\cal D}_j\, {\tilde q}^{(2)}\right|^2 
\pm
\xi \, B^{(0)} \, 
\Big)
\label{bog}
\eea
where $ B^{(0)}  \equiv   \frac{ 1}{ 2}  \epsilon_{ij}  {F}^{(0)}_{ij }$ is the magnetic flux density along the $z$ direction.  The first-order  Bogomolnyi equations  are  obtained by setting to zero  all square bracket terms in Eq.~(\ref{bog}),  that is, all terms except the last, topological invariant, winding-number term.  Their solutions 
can be elegantly expressed in terms of the  moduli matrices   ($z \equiv x + i y$)
\be   q^{(1)} =  S_{n}^{-1}(z,\bar z) \, e^{- \lambda_{1} \psi(z, {\bar z})}  \, H_0^{(n)}(z);   \quad   {\tilde q}^{(2)} =  S_{r}^{-1}(z,\bar z) \,  e^{- \lambda_{2}\psi(z, {\bar z})} \, H_0^{(r)}(z); 
\label{modmatrep}  \ee
where $ H_0^{(n)}(z)$ and $ H_0^{(r)}(z)$ are $n \times n$ and $r \times r$  matrices {\it holomorphic}  in $z$,
while  $S_{n}$ ($S_{r}$) is a regular  $SL(n, C)$   ($SL(r, C)$) matrix; $\psi(z, {\bar z})$ is a complex function, which can be chosen real by an appropriate choice of gauge.   
\be   \lambda_{1}=   \frac{r}{\sqrt{2 \,n\, r\,  ( r + n ) }}, \quad   \lambda_{2}=   \frac{n}{\sqrt{2 \,n\, r\,  ( r + n ) }}\,
\label{thecharges}\ee
are the $U(1)$ charges of the $q^{(1)}$ and   ${\tilde q}^{(2)}$ fields, respectively,  see Eq.~(\ref{U1gene}).  
 $S_{n}$ ($S_{r}$)  corresponds to the complexified $SU(n)$ ($SU(r)$)
transformations.  Note that $H_{0}$'s and $S$'s are defined up to transformations of the form
   \[     H_0^{(n)}(z) \to  V_{n}(z) \,   H_0^{(n)}(z) ; \quad S_{n}(z,\bar z) \to   V_{n}(z)\,S_{n}(z,\bar z)\,,  \]
where  $V_{n}(z)$ is an arbitrary regular, holomorphic  $n \times n$  ({\it vis-\`a-vis},   $r \times r$ for $H_0^{(r)}(z)$, $S_{r}$)
matrix of determinant one. $ H_0^{(n)}(z)$ and $ H_0^{(r)}(z)$, called
moduli matrices, contain all the moduli parameters \cite{Eto:2006pg}. $SU(n)$, $SU(r)$, $U(1)$ gauge fields are given by  ($\bar \partial \equiv  \de / \de {\bar z}$)
\bqa  &&  A^{(n)}_1 + i\,A^{(n)}_2 = - 2\,i\,S_{n}^{-1}(z,\bar z) \, \bar\partial  \, S_{n}(z,\bar z)\,; \quad A^{(r)}_1 + i\,A^{(r)}_2 = - 2\,i\,S_{r}^{-1}(z,\bar z) \, \bar\partial \,  S_{r}(z,\bar z)\,; 
 \non  \\
&&    A^{(0)}_1 + i\,A^{(0)}_2 = - 2 \,i \, \bar\partial \, \psi\;. \label{from} \eea
These Ans\"atze solve the matter part of the Bogomolnyi equations
\be
\left ({\cal D}_1+i {\cal D}_2  \right) \, q^{(1)}  = \left ({\cal D}_1+i {\cal D}_2  \right) \, {\tilde q}^{(2)}  = 0,\label{eq:bps_equation}
\ee
automatically  (they  reduce to ${\bar \partial} H_{0}=0$).  In order to simplify the (linearized) gauge field equations let us introduce
\[   \Omega_{n} = S_{n}\, S_{n}^{\dagger}, \quad  \Omega_{r} = S_{r}\, S_{r}^{\dagger}\;;
\]
the (Bogomolnyi) gauge field equations  (sometimes called {\it master equations}) are\footnote{For instance, the $SU(n)$ gauge field components can be written from Eq.~(\ref{from})  as
\[  A_{1}= -i (S^{-1} \bar {\partial} S +  S^{\dagger} \partial (S^{\dagger})^{-1} )\;; \quad 
 A_{2}= - (S^{-1} \bar {\partial} S  -   S^{\dagger} \partial (S^{\dagger})^{-1} )\;.
\]
By a straightforward algebra one finds then ($F_{12} = \partial_{1} A_{2}-  \partial_{2} A_{1} + i [A_{1}, A_{2}]$):  
\[ (S^{\dagger})^{-1}\, F_{12} \, S^{\dagger} =  -2 \, \partial (  \Omega^{-1} {\bar \partial} \, \Omega )\;, \quad \Omega = S\, S^{\dagger}\;.\]
 } 
 \[    \partial (  \Omega_{n}^{-1} {\bar \partial} \,\Omega_{n} ) =  \frac{g_{n}^{2}}{4} e^{-2 \lambda_{1} \psi}\, 
 [  \, \Omega_{n}^{-1}\, H_{0}^{(n)}  H_{0}^{(n) \,\dagger}  - \frac{1}{n}  \, \Tr_{n}  \, (\Omega_{n}^{-1}\, H_{0}^{(n)}  H_{0}^{(n) \,\dagger} ) \, {\bf 1}_{n \times n}\,]\;; 
 \]
  \[    \partial (  \Omega_{r}^{-1} {\bar \partial} \,\Omega_{r} ) =  \frac{g_{r}^{2}}{4} e^{-2  \lambda_{2} \psi}\, 
 [  \, \Omega_{r}^{-1}\, H_{0}^{(r)}  H_{0}^{(r) \,\dagger}  - \frac{1}{ r}  \, \Tr_{r}  \, (\Omega_{r}^{-1}\, H_{0}^{(r)}  H_{0}^{(r) \,\dagger} )\, {\bf 1}_{r \times r}\,]\;;
 \]
 \[   \partial {\bar \partial}  \, \psi  = \frac {g_{0}^{2}}{4}  \, [\, 
 \lambda_{1}\,e^{-2 \lambda_{1} \psi}\,  \Tr_{n} \, ( \Omega_{n}^{-1}\, H_{0}^{(n)}  H_{0}^{(n)\, \dagger})  +   \lambda_{2}\, e^{-2 \lambda_{2} \psi}\, \Tr_{r} \, ( \Omega_{r}^{-1}\, H_{0}^{(r)}  H_{0}^{(r)\, \dagger})  - \xi \,] \;.   
 \]
Since  $SU(n)$, $SU(r)$ and $U(1)$ all commute with each other,  the above construction is basically just a straightforward generalization of the formulas given in the case of $U(n)\sim SU(n) \times U(1)$  theory, see e.g., \cite{Eto:2005yh}, except for one point.  As there is just one $U(1)$ gauge group factor but two non-Abelian groups
$SU(n)$ and $SU(r)$, the moduli matrices are subject to a constraint.  In fact, from Eq.~(\ref{modmatrep}) and the fact that $S_{n}$ ($S_{r}$)  belongs to  $SL(n, C)$   ($SL(r, C)$) it follows that 
\[   e^{- 2 \, \lambda_{1}\, n \,\psi }\, \det H_{0}^{(n)}  H_{0}^{(n)\, \dagger} =  \det ( q^{(1)} q^{(1)\, \dagger})\;;
\]
 \[   e^{- 2 \, \lambda_{2}\, r \,\psi }\, \det H_{0}^{(r)}  H_{0}^{(r)\, \dagger} = \det (  {\tilde q}^{(2)}  {\tilde q}^{(2)\, \dagger} )\;.
\]
As   $\lambda_{1}\, n =\lambda_{2}\, r$  (see Eq.~(\ref{thecharges})),  these are consistent with the asymptotic behavior, 
\[     q^{(1)} q^{(1)\, \dagger}  \sim   |v_{1}|^{2}  \, {\mathbbm 1}_{n \times n},\qquad 
    {\tilde q}^{(2)}  {\tilde q}^{(2)\, \dagger}  \sim   |v_{2}|^{2}  \, {\mathbbm 1}_{r \times r}\,,
\]
if a constraint 
\be   \frac{\det H_{0}^{(n)}  H_{0}^{(n)\, \dagger} }{\det H_{0}^{(r)}  H_{0}^{(r)\, \dagger}} \sim   \frac{ |v_{1}|^{2 n}  }{|v_{2}|^{2r}}\; 
\label{vincolo}\ee
is satisfied at large $|z|$.  So for a vortex of winding number $k$,  
\[  \det H_{0}^{(n)}  H_{0}^{(n)\, \dagger}  \propto  \det H_{0}^{(r)}  H_{0}^{(r)\, \dagger} \sim |z|^{2k}, 
\]
i.e.,  the same winding in $q$ and ${\tilde q}$ fields,   but with the condition, Eq.~(\ref{vincolo}). 

The tension for the minimum vortex  ($k=1$) can be worked out easily as follows. A typical such vortex has the form, Eq.~(\ref{modmatrep}), where the moduli matrices can  be brought to the form locally, e.g.,  
\[     H_0^{(n)}(z)= \left(\begin{array}{cc} c_{1}\, z & {\bf 0} \\{\bf 0} & {\mathbbm 1}_{(n-1)\times(n-1)}\end{array}\right); \quad 
 H_0^{(r)}(z)= \left(\begin{array}{cc} c_{2}\, z & {\bf 0} \\{\bf 0} & {\mathbbm 1}_{(r-1)\times(r-1)}\end{array}\right)\;,
\]
with
\be    \frac{c_{1}}{c_{2}} = \frac{v_{1}^{n}}{v_{2}^{r}}\;. \label{fixed}
\ee
Note that one of $c_{1}$ and  $c_{2}$, for instance $c_{1}$,  can be set to unity by an appropriate choice of $S_{n}$ and $\psi$.   The other is then fixed uniquely. 
In order for the behavior  (by setting $c_{1}=1$) 
\[   H_0^{(n)}(z) \,H_0^{(n)}(z)^{\dagger}  =   \left(\begin{array}{cc}\rho^2 & {\bf 0} \\{\bf 0} & {\mathbbm 1}_{(n-1)\times(n-1)}\end{array}\right)
\]
to be consistent with  $q^{(1)} q^{(1)\, \dagger}  \sim   |v_{1}|^{2}  \, {\mathbbm 1}_{n \times n}$,  the large $\rho$ behavior of $\psi $ and    $ S_{n}$  must be such that 
\[    e^{- 2 \lambda_{1} \psi } \,S_{n}^{-1}\,  (S_{n}^{\dagger})^{-1} \sim   \left(\begin{array}{cc} {1/\rho^2}  & {\bf 0} \\{\bf 0} & {\mathbbm 1}_{(n-1)\times(n-1)}\end{array}\right);
\]
and this is possible if     
\[    S_{n} \sim      \left(\begin{array}{cc}e^{(n-1)\lambda_1\psi } & {\bf 0} \\{\bf 0} &  e^{-\lambda_1 \psi}\,  {\mathbbm 1}_{(n-1)\times(n-1)}\end{array}\right)
\]   
and 
\[    e^{- 2 \,n \,\lambda_{1}\, \psi}  \sim  1/\rho^{2}\;, \qquad .^{.}. \quad \psi \sim\sqrt{\frac{n+r}{2\,n\, r}}\, \log \rho^{2}\;.  
\]
Of course, the same conclusion for $\psi$ is reached by considering the asymptotic behavior of  ${\tilde q}^{(2)}$ and $S_{r}$.  
As  $ F_{12}^{(0)} = - 4 \,{\bar \partial }\, \partial \, \psi $
 \[ T=  \xi  \, \int d^{2} x  \, F_{12}^{(0)} =  {\xi} \, \int d^{2} x  \, \nabla^{2} \psi =  4\,\pi \, \sqrt{\frac{n+r}{2\,n\, r}}\,  \xi
 = 4\,\pi \, \sqrt{\frac{n+r}{n\,r}}\,  \mu \, m_{0},
\]
  that is 
  \[  T = 4 \, \pi \, (|v^{(1)}|^{2} + |v^{(2)}|^{2})\;.
  \]
 
 An (axially symmetric)  vortex of generic $SU(n)\times SU(r)$  orientations can be represented by the moduli matrix of the form, 
\[     H_0^{(n)}(z)= \left(\begin{array}{cc} c_{1} z & {\bf 0} \\ {\mathbbm b} & {\mathbbm 1}_{(n-1)\times(n-1)}\end{array}\right); \quad 
 H_0^{(r)}(z)= \left(\begin{array}{cc} c_{2}  z & {\bf 0} \\{\mathbbm c} & {\mathbbm 1}_{(r-1)\times(r-1)}\end{array}\right)\;.
\]
 where ${\mathbbm b}$  (${\mathbbm c}$) is an $(n-1)$- component ($(r-1)$- component) complex vector, representing the inhomogeneous coordinates of $CP^{n-1}$  (of $CP^{r-1}$).  Under the color-flavor $SU(n)$ ($SU(r)$)  symmetry group
 they transform as  in the fundamental representation of $SU(n)$ ($SU(r)$).  This is the content of some of the claims made in the main text.  
 
 The BPS equations actually allow more general kinds of vortex solutions. The moduli space,  for general winding numbers and with more general position and orientation parameters, shows a very rich and interesting spectrum. 
 This and other questions will be discussed elsewhere.


\begin{thebibliography}{122}

\bibitem{HT}
A.~Hanany and D.~Tong,
JHEP {\bf 0307} (2003) 037
[arXiv:hep-th/0306150].


\bibitem{ABEKY}
R.~Auzzi, S.~Bolognesi, J.~Evslin, K.~Konishi and A.~Yung,
Nucl.\ Phys.\ B {\bf 673} (2003) 187
[arXiv:hep-th/0307287].

\bibitem{ABEK}
  R. Auzzi, S. Bolognesi, J. Evslin,  K. Konishi,
  Nucl. Phys. B {\bf 686}, 119  (2004). 
   [arXiv:hep-th/0312233].

 \bibitem{SY}
  M. Shifman and A. Yung,  
  Phys. Rev.  D {\bf 70},  045004  (2004).  
  [arXiv:hep-th/0403149].
 
 \bibitem{HT2}
A.~Hanany and D.~Tong, 
JHEP {\bf 0404} (2004) 066. 
[arXiv:hep-th/0403158].

\bibitem{Eto:2004rz}
  M. Eto, Y. Isozumi, M. Nitta, K. Ohashi,  N. Sakai,
  Phys. Rev. D  {\bf 72}, 025011  (2005).
     [arXiv:hep-th/0412048]. 
     
     \bibitem{GSY}
A. Gorsky,  M. Shifman,   A. Yung, 
Phys. Rev. D  {\bf 71},    045010 (2005).
   [arXiv:hep-th/0412082].
   
 \bibitem{HTong}  
K. Hashimoto,  D.  Tong,
JCAP {\bf 0509}: 004,2005. 
   [arXiv:hep-th/0506022]. 
   
\bibitem{SB1}     
S.  Bolognesi, 
Nucl. Phys. {\bf B730}: 127-149 (2005). 
   [arXiv:hep-th/0507273].
   

\bibitem{Tong:2005un}
D.~Tong,
``TASI lectures on solitons,''
[arXiv:hep-th/0509216].
   
\bibitem{Eto:2005yh}
M.~Eto, Y.~Isozumi, M.~Nitta, K.~Ohashi and N.~Sakai,
Phys.\ Rev.\ Lett.\  {\bf 96} (2006) 161601
[arXiv:hep-th/0511088].

\bibitem{ASY}
 R. Auzzi, M. Shifman, A. Yung,
 Phys. Rev. {\bf D73}:105012 (2006),  Erratum-ibid. {\bf D76}:109901 (2007). 
[arXiv:hep-th/0511150]. 

\bibitem{SB2}
S. Bolognesi,
  Nucl. Phys. {\bf B752}: 93-123, (2006). 
[arXiv:hep-th/0512133].

\bibitem{AuzziShif2}
R. Auzzi, M. Shifman, A. Yung, 
  Phys. Rev. D {\bf 74}:045007 (2006). 
[arXiv:hep-th/0606060]. 

\bibitem{Eto:2006pg}
M.~Eto, Y.~Isozumi, M.~Nitta, K.~Ohashi and N.~Sakai,
J.\ Phys.\ A {\bf 39} (2006) R315
[arXiv:hep-th/0602170].

\bibitem{Shifman:2006kd}
M.~Shifman and A.~Yung,
Phys.\ Rev.\ D {\bf 73} (2006) 125012
[arXiv:hep-th/0603134].

\bibitem{Eto:2006cx}
M.~Eto, K.~Konishi, G.~Marmorini, M.~Nitta, K.~Ohashi, W.~Vinci and N.~Yokoi,
Phys.\ Rev.\ D {\bf 74}, 065021 (2006)
[arXiv:hep-th/0607070].

\bibitem{Eto:2006db}
  M.~Eto, K.~Hashimoto, G.~Marmorini, M.~Nitta, K.~Ohashi and W.~Vinci,
  Phys.\ Rev.\ Lett.\  {\bf 98} (2007) 091602
  [arXiv:hep-th/0609214].

\bibitem{Eto:2006dx}
M.~Eto, L.~Ferretti, K.~Konishi, G.~Marmorini, M.~Nitta, K.~Ohashi, W.~Vinci and N.~Yokoi,
Nucl. Phys. {\bf B780}:161-187 (2007)  [arXiv:hep-th/0611313].

\bibitem{TongHetero}
M. Edalati, D. Tong, 
 JHEP {\bf 0705}:005,2007. 
  [arXiv:hep-th/0703045].

\bibitem{TongHetero2}
David Tong (Cambridge U., DAMTP) . Mar 2007. 13pp. 
JHEP {\bf 0709}:022, 2007. 
  [arXiv:hep-th/0703235]. 

\bibitem{SYReview} 
M. Shifman, A. Yung,
``Supersymmetric Solitons and How They Help Us Understand Non-Abelian Gauge Theories'',  Rev.Mod.Phys. {\bf 79}:1139 (2007)
  [arXiv:hep-th/0703267]. 

\bibitem{SemiSeven}
M.~Eto, J.~Evslin, L.~Ferretti, K.~Konishi, G.~Marmorini, M.~Nitta, K.~Ohashi, W.~Vinci and N.~Yokoi,
Phys.\ Rev.\ D {\bf 76} 105002 (2007),
  arXiv:0704.2218 [hep-th].

\bibitem{GFK}
L.  Ferretti, S. B. Gudnason, K. Konishi, 
 Nucl. Phys. {\bf B789} 84-110 (2008) 
arXiv:0706.3854 [hep-th].

\bibitem{EHT}
M. Eto, K. Hashimoto, S. Terashima, JHEP {\bf 0709}:036,2007.
arXiv:0706.2005 [hep-th].


\bibitem{SW1}
N. Seiberg,  E. Witten,
 Nucl. Phys.   B {\bf 426},  19   (1994); 
 Erratum
 \textit{ibid.}   B {\bf 430},  485  (1994)
[arXiv:hep-th/9407087].


\bibitem{SW2}
N. Seiberg,  E. Witten,
  Nucl. Phys.   B {\bf 431},  484  (1994)
[arXiv:hep-th/9408099].

\bibitem{curves}
P. C. Argyres,  A. F. Faraggi,    Phys. Rev. Lett {\bf 74}, 3931   (1995), [arXiv:hep-th/9411057]; 
A. Klemm, W. Lerche, S. Theisen, S. Yankielowicz,    Phys. Lett. B 
{\bf 344}, 169     (1995);   Int. J. Mod. Phys.  A {\bf 11}, 1929    (1996), [arXiv:hep-th/9411048], 
A. Hanany,  Y. Oz,   Nucl. Phys. B {\bf 452}, 283  (1995), [arXiv:hep-th/9505075]. 


\bibitem{ND}
N.~Dorey, 
 JHEP {\bf 9811} 005 (1998) [arXiv:hep-th/9806056];
 N. Dorey, T. J. Hollowood and D. Tong,
JHEP {\bf  9905} 006 (1999) [arXiv:hep-th/9902134].
 
\bibitem{Witten} 
E.~Witten, Nucl. Phys. {\bf B188}, 513 (1981). 

\bibitem{CFIV}
S. Cecotti, P. Fendley, K. A. Intriligator and C. Vafa,
Nucl. Phys. {\bf B 386} 405 (1992) [arXiv:hep-th/9204102].


\bibitem{CKM}
G. Carlino, K. Konishi and H. Murayama,
   {\bf  JHEP   0002}  (2000) 004,       [arXiv:hep-th/0001036];
 {\bf    Nucl. Phys.  B590}  (2000) 37,       [arXiv:hep-th/0005076]. 

\bibitem{BKM}   S. Bolognesi, K. Konishi, G. Marmorini,  
Nucl. Phys. B  {\bf 718}, 134  (2005),   [arXiv:hep-th/0502004].

\bibitem{MKY}
 G. Marmorini, K. Konishi,  N. Yokoi,  Nucl. Phys.  B {\bf 741},  180 (2006) [arXiv:hep-th/0502004].  
 
 
 \bibitem{Konishi} 
 ``The Magnetic Monopoles Seventy-Five Years Later'', 
 Lecture Notes in Physics 737 (2007), p.471 (Springer). 
  [arXiv:hep-th/0702102]. 
 

\bibitem{APS}
P. C. Argyres, M. R. Plesser,  N. Seiberg, Nucl. Phys.  B {\bf 471}, 159  
(1996);   
P.C. Argyres, M.R. Plesser,  A.D. Shapere, 
Nucl. Phys. B {\bf 483}, 172 (1997);
 K.  Hori, H. Ooguri,   Y.  Oz,
 Adv. Theor. Math. Phys.   {\bf 1}, 1  (1998).
 
 
 \bibitem{Tumbling}  S. Raby, S. Dimopoulos and  L.  Susskind,  Nucl.Phys. {\bf B169}:373 (1980).
 
 \bibitem{Kogut} 
J. B Kogut,    Rev. Mod. Phy. {\bf 51 } 659 (1979) and references therein. 
 

\bibitem{BPS}  
E.B. Bogomolnyi, Sov. J. Nucl. Phys. {\bf  24}, 449 (1976).


\bibitem{Strassler}
M. J.~Strassler,
 {  JHEP  9809}  (1998)  017
[arXiv:hep-th/9709081], ``On Phases of Gauge Theories and the Role of Non-BPS Solitons in Field Theory '',
III Workshop ``Continuous Advance in QCD'', Univ. of Minnesota  (1998), [arXiv:hep-th/9808073].

\bibitem{JR}  R. Jackiw,  C. Rebbi, 
Phys. Rev. D {\bf 13}, 3398 (1976).  

\bibitem{BK}   S. Bolognesi, K. Konishi,   Nucl. Phys. B {\bf  645},  337 (2002) [arXiv:hep-th/0207161].

\bibitem{AGK} R. Auzzi, R. Grena,  K. Konishi,    Nucl. Phys. B {\bf 653},  204  (2003) [arXiv:hep-th/0211282]. 


\bibitem{Kanomaly}  K. Konishi,   Phys.Lett. {\bf B135}: 439 (1984).

\bibitem{SYLues} M. Shifman and  A. Yung, Phys.Rev. {\bf D77}: 066008  (2008)
arXiv:0712.3512 [hep-th]. 


\bibitem{SYbook}  M. Shifman and  A. Yung, private communication, to appear. 


\end{thebibliography}
 \end{document}